\documentclass[aps,prd,amsfonts,amssymb, amsmath, showpacs, showkeys, a4paper, nofootinbib, superscriptaddress, 11pt, raggedbottom]{revtex4-2}

\usepackage{braket}
\usepackage{bm}
\usepackage{graphicx}
\usepackage{slashed}
\usepackage{subfig}
\usepackage{mathrsfs}
\usepackage{color}
\usepackage{mathtools}
\usepackage[normalem]{ulem}
\usepackage{slashed}

\newcommand{\cannex}{{\sc{Cannex}}}   

\renewcommand{\vec}[1]{\bm{#1}}

\begin{document}

\title{\Large Quantum and thermal pressures from light scalar fields}

\author{Hauke Fischer}
\email{hauke.fischer@tuwien.ac.at}
\affiliation{Atominstitut, Technische Universit\"at Wien, Stadionallee 2, 1020 Vienna, Austria}

\author{Christian K\"{a}ding}
\email{christian.kaeding@tuwien.ac.at}
\affiliation{Atominstitut, Technische Universit\"at Wien, Stadionallee 2, 1020 Vienna, Austria}

\author{Mario Pitschmann}
\email{mario.pitschmann@tuwien.ac.at}
\affiliation{Atominstitut, Technische Universit\"at Wien, Stadionallee 2, 1020 Vienna, Austria}

\begin{abstract}
Light scalar fields play a variety of roles in modern physics, especially in cosmology and modified theories of gravity. For this reason, there is a zoo of experiments actively trying to find evidence for many scalar field models that have been proposed in theoretical considerations. Among those are setups in which the pressures expected to be induced by light scalar fields between two parallel plates are studied, for example, Casimir force experiments. While it is known that classical and quantum pressures caused by light scalar fields could have significant impacts on such experiments, in this article, we show that this can also be the case for thermal pressure. More specifically, we derive expressions for the quantum and thermal pressures induced by exchanges of light scalar field fluctuations between two thin parallel plates. As particular examples, we then look at screened scalar fields. For chameleon, symmetron and environment-dependent dilaton models, we find large regions in their parameter spaces that allow for thermal pressures to equal or exceed the quantum pressures. By comparing with earlier constraints from quantum pressure calculations, we conclude that thermal pressures induced by chameleons are actually of experimental significance.
\end{abstract}


\maketitle


\section{Introduction}

Light scalar fields appear in many contexts in modern cosmology or modifications of general relativity, for example, in scalar-tensor theories of gravity \cite{Fujii2003}. They often serve as the basis for proposed solutions of some of modern physics' greatest open problems, including dark energy (DE) and dark matter (DM) \cite{Clifton2011,Joyce2014}. Many of such theories give rise to a universal coupling between the scalar degree of freedom and Standard Model matter, leading to an additional force of Nature. Such a fifth force, however, is tightly constrained within our Solar System \cite{Dickey1994,Adelberger2003,Kapner2007}. 

Screening mechanisms were devised to provide a way of circumventing these constraints while also offering a rich phenomenology. They strongly suppress (screen) scalar fifth forces in environments of sufficiently high mass densities, e.g., our Solar System, but enable them to develop their full strengths in regions of lower densities. There are a variety of different light scalar field models with screening mechanisms, so-called screened scalar fields, some of which are even being discussed as presenting an alternative to particle DM through their fifth forces \cite{Burrage:2016yjm,OHare:2018ayv,Burrage:2018zuj,Kading:2023hdb}. Some of the most popular screened scalar field models are chameleons~\cite{Khoury2003,Khoury20032}, symmetrons~\cite{Dehnen1992, Gessner1992, Damour1994, Pietroni2005, Olive2008, Brax:2010gi,Hinterbichler2010,Hinterbichler2011}, and environment-dependent dilatons~\cite{Damour1994,Gasperini:2001pc,Damour:2002nv,Damour:2002mi,Brax:2010gi,Brax:2011ja,Brax2022}. They are the examples used in the present article, and have already been tested and discussed in multiple experimental setups; see, e.g., Refs.~\cite{Burrage:2016bwy,Pokotilovski:2012xuk,Pokotilovski:2013tma,Jenke:2014yel,Baum:2014rka,Burrage:2014oza,Hamilton:2015zga,Lemmel:2015kwa,Burrage:2015lya,Elder:2016yxm,Ivanov:2016rfs,Burrage:2016rkv,Jaffe:2016fsh,Brax:2017hna,Sabulsky:2018jma,Brax:2018iyo,Cronenberg:2018qxf,Hartley:2018lvm,Zhang:2019hos,ArguedasCuendis:2019fxj,Jenke:2020obe,Pitschmann:2020ejb,Brax:2021wcv,Qvarfort:2021zrl,Yin:2022geb,Betz:2022djh, Brax:2022olf, Hartley:2019wzu,Fischer:2023koa,Fischer:2023eww,Fischer:2024coj,Klimchitskaya:2024dvk,Rene,Baez-Camargo:2024jia,OShea:2024jjw,Mueller2024}. For an overview of current constraints on the parameter spaces of these models, see Refs.~\cite{Fischer:2024eic,Burrage:2017qrf}. While screened scalar fields are usually treated as classical fields, there have already been first attempts to describe them in quantum field theoretical frameworks \cite{Brax:2018grq,Burrage:2018pyg,Burrage:2019szw,Kading:2023mdk} and from the perspective of particle physics \cite{SevillanoMunoz:2024ayh}. The present article will add to this. 

Casimir force experiments like the upcoming \textit{Casimir And Non Newtonian force EXperiment} (\cannex{})~\cite{Sedmik:2021iaw,Rene} are expected to further constrain the parameter spaces of light scalar field models \cite{Brax:2014zta,Almasi:2015zpa,Elder:2019yyp,Brax:2022uiv,Fischer:2023koa}. This makes this type of experiments an interesting subject for theoretical analyses and motivates the investigation in the present article. Ref.~\cite{Brax:2018grq} initiated the discussion of quantum pressure induced between two parallel plates, as can be found in \cannex{}, by interactions with quantum fluctuations of screened scalar fields. For this, the plates were approximated by having infinite widths and infinite thicknesses. Using this approximation, Ref.~\cite{Brax:2018grq} showed that the quantum pressures of chameleons and symmetrons can actually be quite significant and lead to tighter constraints on these models' parameter spaces. In the present article, we consider two equal, infinitely wide, but very thin plates and also study the quantum pressures induced by light scalar fields, while considering chameleons, symmetrons, and environment-dependent dilatons as representative examples. In order to obtain analytically tractable results, we restrict our analysis to parameter regimes in which these screened scalar fields have Compton wavelengths much larger than the thickness of one plate. This means that throughout our investigation, we can safely assume the scalar fields to have constant masses defined by the density of the vacuum surrounding the plates. Consequently, the general parts of our results are applicable to any type of scalar field model that exhibits a constant mass in the considered situation. Motivated by the results of Refs.~\cite{Burrage:2018pyg,Burrage:2019szw,Kading:2023mdk}, which indicated that finite-temperature effects of light scalar fields could be relevant for experiments, we also compute expressions for the pressures induced by thermal scalar fluctuations. Comparing the results for quantum and thermal pressures, we find that both can be of comparable magnitudes in some situations. Due to the proven significance of quantum pressures according to Ref.~\cite{Brax:2018grq}, we consequently conclude that thermal pressures are likely also relevant for screened scalar field models in some experimental setups and must therefore be taken into account. 

This article is structured as follows: At first, in Sec.~\ref{sec:Setup}, we shortly introduce the three considered screened scalar field models and discuss their couplings to fermions. Next, in Sec.~\ref{sec:QTpress}, we compute the quantum and thermal potentials, and derive and discuss the resulting pressures between two parallel plates. Finally, we draw our conclusions in Sec.~\ref{sec:Conclusion}.


\section{Scalar field models}
\label{sec:Setup}

While it should be noted that some of the main results in this article are applicable to more general types of light scalar fields, we will use three popular types of screened scalar fields models as examples for our discussion. In this way, we are able to make comparisons with the results of Ref.~\cite{Brax:2018grq}. Therefore, in this section, we first give a short overview of the considered screened scalar field models and provide all necessary formulas. Afterwards, we discuss the couplings of the scalar fields to fermions since they are required for the computations done in this article.


\subsection{Models}

As examples, we consider three screened scalar field models: chameleons, symmetrons, and environment-dependent dilatons. Each of them can be generally described by the Einstein frame action 
\begin{eqnarray}\label{eq:Action}
S &=& \int  d^4x\, \sqrt{-g} \left( -\frac{m_\text{Pl}^2}{2}\,R + \frac{1}{2}(\partial\varphi)^2 - V_X(\varphi) \right) + \int d^4x\,\sqrt{-\tilde g^X}\,\mathcal{L}_\text{SM} (\tilde g_{\mu \nu}^X,\psi_i)~,
\end{eqnarray}
in which $m_\text{Pl}$ is the reduced Planck mass; $\varphi$ represents the scalar field; $X \in \{C,S,D \}$ labels the scalar field model, i.e., chameleon ($C$), symmetron ($S$), or dilaton ($D$); $V_X(\varphi)$ is the scalar's potential; and $\mathcal{L}_\text{SM}$ denotes a Lagrangian describing the Standard Model fields $\psi_i$. For notational convenience, in Eq.~(\ref{eq:Action}), we kept the Jordan frame metric defined in terms of the conformal factor $A_X(\varphi)$ as $\tilde g^X_{\mu \nu} = A_X^2(\varphi)g_{\mu \nu}$. Due to the conformal factor, the scalar $\varphi$ couples to the field $\psi_i$. In turn, this leads to an effective potential for the scalar:
\begin{eqnarray}\label{eq:EffPot}
    V_{X;\text{eff}} (\varphi) &:=& V_X(\varphi)+ A_X(\varphi)T^\mu_{\phantom{\mu}\mu}
\end{eqnarray}
with $T^\mu_{\phantom{\mu}\mu}$ as the trace of the energy momentum tensor of $\psi_i$. 

For chameleons, we have 
\begin{eqnarray}\label{eq:Cham}
    V_C &=& \frac{\Lambda^{n+4}}{\varphi^n}~,~A_C(\varphi) \,=\, \exp\left(\frac{\varphi}{M_C}\right)\,\approx\, 1 + \frac{\varphi}{M_C}+ \frac{\varphi^2}{2M_C^2}~,
\end{eqnarray}
where $n \in \mathbb{Z}^+ \cup 2\mathbb{Z}^-\setminus\{-2\}$ defines the exact chameleon model; $\Lambda$ is a mass scale determining the chameleon's self-interaction, except for the case $n=-4$, in which case it must actually be a dimensionless constant; and $M_C$ is another mass scale, which controls the chameleon coupling to matter. Here, we have assumed $\varphi \ll M_C$ and kept the second order term in order to allow for comparisons with symmetron and dilaton models. If we only consider the conformal factor up to the first order in $\varphi/M_C$, the effective potential defined by Eqs.~(\ref{eq:EffPot}) and (\ref{eq:Cham}) gives rise to an effective chameleon mass
\begin{eqnarray}
\label{eq:CMass}
    m_C^2 &=& \frac{n(n+1)\Lambda^{n+4}}{\varphi_C^{n+2}}
\end{eqnarray}
with the chameleon vacuum expectation value (VEV) given by 
\begin{eqnarray}
\label{eq:CVEV}
    \varphi_C &=& \left( n \Lambda^{n+4} \frac{M_C}{T^\mu_{\phantom{\mu}\mu}} \right)^{1/(n+1)}~.
\end{eqnarray}
While we have kept the second order term of the conformal factor in Eq.~(\ref{eq:Cham}) in order to enable us to discuss the exchange of two chameleon field fluctuations, the second order corrections to Eqs.~(\ref{eq:CMass}) and (\ref{eq:CVEV}) are not relevant in the following discussions. Clearly, for any permitted value of $n$, a larger $T^\mu_{\phantom{\mu}\mu}$ implies a larger $m_C$ in Eq.~(\ref{eq:CMass}). Considering the example of non-relativistic matter, this means that a quantum of a chameleon field is the heavier the denser its surrounding environment it couples to. A heavier chameleon induces a shorter-ranged fifth force. This is the essence of the chameleon screening mechanism \cite{Khoury2003,Khoury20032}.

The symmetron model is defined by
\begin{eqnarray}\label{eq:Symm}
    V_S &=& -\frac{\mu^2}{2}\,\varphi^2 + \frac{\lambda_S}{4}\,\varphi^4~,~A_S(\varphi) \,=\, 1 + \frac{\varphi^2}{2M_S^2}
\end{eqnarray}
with the tachyonic mass $\mu$, the dimensionless self-coupling parameter $\lambda_S$, and the mass scale $M_S$ defining the symmetron's coupling to matter. It is screened by the Damour-Polyakov mechanism \cite{Damour1994}, which means that the symmetron's effective coupling to matter varies with the density of the environment. In the case $T^\mu_{\phantom{\mu}\mu} \geq \mu^2 M_S^2$, the symmetron VEV must vanish, which results in a complete decoupling from matter if quantum or thermal fluctuations of the symmetron are small enough to be ignored. However, if $T^\mu_{\phantom{\mu}\mu} < \mu^2 M_S^2$, the symmetron has a non-vanishing VEV
\begin{eqnarray}
    \varphi_S &=& \pm \sqrt{\frac{1}{\lambda_S} \left(\mu^2 - \frac{T^\mu_{\phantom{\mu}\mu}}{M_S^2}\right) }~,
\end{eqnarray}
which implies that the symmetron fifth force is unscreened. The symmetron mass is given by
\begin{eqnarray}
\label{eq:Symmmass}
    m_S^2 &=& 
    \begin{cases}
    \frac{T^\mu_{\phantom{\mu}\mu}}{\mathcal{M}^2}-\mu^2 &,~ T^\mu_{\phantom{\mu}\mu} \geq \mu^2 M_S^2
    \\
    2\left(\mu^2-\frac{T^\mu_{\phantom{\mu}\mu}}{\mathcal{M}^2} \right)&,~ T^\mu_{\phantom{\mu}\mu} < \mu^2 M_S^2
    \end{cases}~.
\end{eqnarray}
Note that the symmetron also has an environment-dependent mass and is therefore, technically, a chameleon as well. However, the variation of the effective mass usually has only little influence on the fifth force screening of this model.

Finally, the environment-dependent dilaton has 
\begin{eqnarray}
    V_D &=& V_0\, {e}^{-\lambda \varphi /m_{\text{Pl}}}~,~A_D(\varphi) \,=\, 1 + A_2\,\frac{\varphi^2}{2m_{\text{Pl}}^2}~,
\end{eqnarray}
where $V_0$ is a constant energy density, $\lambda$ denotes the dilaton's dimensionless self-coupling constant, and $A_2$ is a dimensionless constant determining the coupling to matter. Resulting from this, the dilaton has a VEV \cite{Brax2022}
\begin{eqnarray}\label{eq:DilVEV}
  \varphi_D &=& \frac{m_\text{Pl}}{\lambda}\,W\bigg(\frac{\lambda^2V_0}{A_2T^\mu_{\phantom{\mu}\mu}}\bigg)
\end{eqnarray}
with the Lambert $W$-function
\begin{eqnarray}
   W(x) &=& \sum_{n=1}^\infty\frac{(-n)^{n-1}}{n!}\,x^n ~.
\end{eqnarray}
Consequently, its mass is given by 
\begin{eqnarray}
   m_D^2 &=& \frac{1}{m_\text{Pl}^2}\left(\lambda^2V_0\,e^{-\lambda\varphi_D/m_\text{Pl}} + A_2T^\mu_{\phantom{\mu}\mu}\right)~.
\end{eqnarray}
As can be inferred from Eq.~(\ref{eq:DilVEV}), the dilaton VEV is the smaller the larger $T^\mu_{\phantom{\mu}\mu}$. This means that, as long as any dilaton fluctuations are negligible, the coupling between dilaton and matter is small in dense environments. In turn, this leads to the dilaton fifth being rendered weak in such environments. Initially, it was thought that the dilaton was only screened by this realisation of the Damour-Polyakov mechanism. However, Ref.~\cite{Fischer:2023koa} showed that this is only partially true. In fact, there are only certain parts of the dilaton's parameter space in which the Damour-Polyakov mechanism is the dominant reason for fifth force screening. In other parameter regions, the dilaton fith force is mainly screened by the chameleon mechanism. 


\subsection{Couplings to fermions}

For the computations in this article, we will consider that the screened scalar fields couple to the nucleons in the two parallel plates. Consequently, we must describe the interactions between screened scalar fields and fermions.

From the effective actions of the considered models, we can conclude for the following interaction Lagrangians:
\begin{eqnarray}
\label{eq:Lagrangians}
    \mathcal{L}_{C;\text{int}} &=& - \left( \frac{\varphi}{M_C} +  \frac{\varphi^2}{2M_C^2}\right)T^\mu_{\phantom{\mu}\mu}
    ~,~
    \mathcal{L}_{S;\text{int}} \,=\, -  \frac{\varphi^2}{2M_S^2}T^\mu_{\phantom{\mu}\mu}
    ~,~
    \mathcal{L}_{D;\text{int}} \,=\, - A_2 \frac{\varphi^2}{2m^2_{\text{Pl}}}T^\mu_{\phantom{\mu}\mu}~.
\end{eqnarray}
We consider an interaction with a Dirac particle of mass $m$ with stress-energy tensor
\begin{eqnarray}
    T^\mu_{\phantom{\mu}\nu} &=& \bar{\psi} \mathrm{i} \gamma^\mu \partial_\nu \psi - \bar{\psi} \mathrm{i} \slashed{\partial}\psi \delta^\mu_{\phantom{\mu}\nu} + m \bar{\psi}\psi \delta^\mu_{\phantom{\mu}\nu}~.
\end{eqnarray}
From this follows the trace
\begin{eqnarray}
    T^\mu_{\phantom{\mu}\mu} &=&  - 3\bar{\psi} \mathrm{i} \slashed{\partial}\psi + 4m \bar{\psi}\psi 
    \nonumber
    \\
    &=& m\bar{\psi}\psi~,
\end{eqnarray}
where we have used the Dirac equation in the last step. After expanding the fields in terms of their VEVs and fluctuations $\phi$, such that $\varphi = \varphi_X + \phi$, we find that the Lagrangians in Eq.~(\ref{eq:Lagrangians}) give rise to interactions of fermions with one or two scalar fluctuations. This can be summarised in terms of two general interaction Lagrangians:
\begin{eqnarray}
\label{eq:L1}
\mathcal{L}_{X;\text{int}}^{(1)} &=& - g_X^{(1)} \phi\bar{\psi}\psi
~,~
\mathcal{L}_{X;\text{int}}^{(2)} \,=\, -g_X^{(2)}\phi^2 \bar{\psi}\psi~.
\end{eqnarray}
The dimensionless $g_X^{(1)}$ are given by
\begin{eqnarray}
\label{eq:g1}
    g_C^{(1)} &\approx& \frac{m}{M_C}
    ~,~
    g_S^{(1)} \,=\, \frac{\varphi_S m}{M_S^2}
    ~,~
    g_D^{(1)} \,=\, \frac{A_2\varphi_D m}{m_{\text{Pl}}^2}~,
\end{eqnarray}
while the $g_X^{(2)}$ have dimensions of an inverse mass and are
\begin{eqnarray}
\label{eq:g2}
    g_C^{(2)} &=& \frac{m}{2M_C^2}
    ~,~
    g_S^{(2)} \,=\, \frac{m}{2M_S^2}
    ~,~
    g_D^{(2)} \,=\, \frac{A_2 m}{2m_{\text{Pl}}^2}~.
\end{eqnarray}
In Eq.~(\ref{eq:g1}), we have dropped the second order term of $g_C^{(1)}$ since it will always be much smaller than the first order term. However, in order to also allow for a two-scalar exchange, we have kept the second order term that defines $g_C^{(2)}$ in Eq.~(\ref{eq:g2}).


\section{Quantum and thermal pressures}
\label{sec:QTpress}

In this section, we compute the quantum and thermal pressures induced by light scalar fields between two infinitely wide, but very thin plates. For this, we first derive the quantum and thermal potentials from the interaction Lagrangians in Eq.~(\ref{eq:L1}). From those we then obtain the corresponding pressures. Finally, we compare quantum and thermal pressures in order to identify conditions for which both are of comparable magnitude. The obtained results are generally applicable to any scalar field model that has, in the considered experimental situation, a constant mass and couples to fermions as in Eq.~(\ref{eq:L1}).

Note that, for our examples of screened scalar fields, we assume a constant mass $m_X$ throughout this discussion. This mass is defined only by the constant density of the vacuum or gas surrounding the two parallel plates. Such an assumption is well-justified as long as the Compton wavelength of a screened scalar field within one of the plates is much larger than the plate's thickness. In this case, the scalar field will not be able to minimize its potential within the plate. Consequently, the density of the plate will only lead to small perturbations of the scalar's VEV and mass between the plates, which we can safely ignore. Certainly, this restricts the validity of the following discussion only to regions of the model parameter spaces that can comply with the requirements of our assumption.


\subsection{Potentials}

We have to consider the exchanges of one or two scalar fluctuations between two nucleons, one within one plate and one within the other plate. At first, we look at the single-particle exchange. In order to compute the quantum potential induced by light scalar fields from the interaction Lagrangian $\mathcal{L}_X^{(1)}$ in Eq.~(\ref{eq:L1}), we follow Ref.~\cite{MarioHabil}. Eq.~(\ref{eq:L1}) implies the scattering amplitude
\begin{eqnarray}
\label{eq:scattamp}
   \mathrm{i}\mathcal{M}_{X;Q}^{(1)} &=& \left(\mathrm{i}g_X^{(1)}\right)^2\,\bar u(p_2',\sigma_2)u(p_2,\sigma_2)\,\frac{\mathrm{i}}{q^2 - m_X^2}\,\bar u(p_1',\sigma_1)u(p_1,\sigma_1)~,
\end{eqnarray}
where the subscript $Q$ indicates that we refer to the amplitude resulting from the quantum fluctuations of the scalar fields.
In the non-relativistic limit, the amplitude takes the form
\begin{eqnarray}
   \mathcal{M}_{X;Q}^{(1)} &=& -\left(\mathrm{i}g_X^{(1)}\right)^2\,2m\,\frac{1}{\vec q^2 + m_X^2}\,2m~. 
\end{eqnarray}
A comparison with
\begin{eqnarray}
   \mathcal{M}_{X;Q}^{(1)} &=& -(2m)^2\,V^{(1)}_{X;Q}(\vec q)~, 
\end{eqnarray}
where $V^{(1)}_{X;Q}(\vec q)$ is the Fourier transformed of the quantum potential $V^{(1)}_{X;Q}(r)$, yields immediately
\begin{eqnarray}
   V^{(1)}_{X;Q}(\vec q) &=& -\left(g_X^{(1)}\right)^2\frac{1}{\vec q^2 + m_X^2}~.
\end{eqnarray}
With the inverse Fourier transform
\begin{eqnarray}
   V^{(1)}_{X;Q}(r) &=& \int\frac{d^3q}{(2\pi)^3}\,V^{(1)}_{X;Q}(\vec q)\,e^{\mathrm{i}\vec q\cdot\vec r}~, 
\end{eqnarray}
we obtain the quantum potential for the single-scalar exchange:
\begin{eqnarray}
\label{eq:Q1}
   V^{(1)}_{X;Q}(r) &=& -\left(g_X^{(1)}\right)^2\frac{1}{4\pi r}\,e^{-m_X r}~.
\end{eqnarray}

Next, we say that the thin parallel plates and the nucleons within them have a constant temperature $T$. Further, we follow Refs.~\cite{Burrage:2018pyg,Burrage:2019szw,Kading:2023mdk} in assuming that the screened scalar field had sufficient time to thermalize with the plates, such that we can expect thermal scalar fluctuations associated with the temperature $T$. In this case, we can add a finite temperature contribution \cite{Ferrer:1998rw} to the propagator. However, as is explained below, the thermal contribution at tree level vanishes for kinematical reasons. In fact, the amplitude for the exchange of a single thermal scalar fluctuation between two nucleons is given by 
\begin{eqnarray}
   \mathrm{i}\mathcal{M}_{X;T}^{(1)} &=& \left(\mathrm{i}g_X^{(1)}\right)^2\,\bar u(p_2',\sigma_2)u(p_2,\sigma_2)\,2\pi \delta(q^2-m_X^2) n(T,E_q)\,\bar u(p_1',\sigma_1)u(p_1,\sigma_1)~,
\end{eqnarray}
where we use the Boltzmann distribution
\begin{eqnarray}
    n(T,E_q) &=& \exp\left( -E_q /T\right)
\end{eqnarray}
with energy $E_q$\footnote{ Note that we have approximated the Bose-Einstein distribution by the Boltzmann distribution for analytical reasons. This restricts our discussion to parts of the scalar field parameter spaces for which $m_X \gg T$.}. Though, since in the rest frame of a nucleon, i.e., $p_1 = (m,\vec{0})^T$, we have
\begin{eqnarray}
    q^2-m_X^2 &=& \left(p_1 - p_1'\right)^2-m_X^2 \nonumber\\
    &=&2m^2 - 2p_1\cdot p_1'-m_X^2 \nonumber\\
    &=&2m\left(m - \sqrt{m^2 + {\vec{p_1}'}^2}\right)-m_X^2 < 0~,
\end{eqnarray}
relativistic invariance implies $\delta(q^2-m_X^2)=0$ as well as $\mathcal{M}_{X;T}^{(1)}=0$, as expected. This means that there are no thermal potential and pressure from a single-scalar exchange between two on-shell fermions.


For the quantum and thermal potentials from the two-scalar exchange, we take and validate the results from Ref.~\cite{Ferrer:1998rw}, which itself makes use of Refs.~\cite{Feinberg:1989ps,Ferrer:1998ju,Horowitz:1993kw}. Therefore, for the quantum contribution, we find 
\begin{eqnarray}
\label{eq:Q2}
    V^{(2)}_{X;Q}(r) &=& - \left(g_X^{(2)}\right)^2\frac{ m_X }{8\pi^3  r^2} K_1(2 m_X r)~.
\end{eqnarray}
Furthermore, for the thermal contribution, we obtain
\begin{eqnarray}
\label{eq:T2}
    V^{(2)}_{X;T}(r) &=& -\left(g_X^{(2)}\right)^2\frac{1}{2\pi^3 } \frac{1}{r} 
    \frac{T m_X}{\sqrt{1+(2rT)^2}} K_1\left( \frac{m_X}{T}\sqrt{1+(2rT)^2} \right)~.
\end{eqnarray}
It should be noted that this usual procedure of obtaining potentials implicitly assumes the nucleons to be on-shell, as is the case for free asymptotic scattering states. However, in the next section, we consider the case in which the nucleons are bound within the macroscopic plates. Hence, for our discussion, such a procedure of obtaining effective macroscopic potentials is an approximation.
 

\subsection{Pressures}

\begin{figure} [htbp]
\centering
    \includegraphics[width=7.8cm]{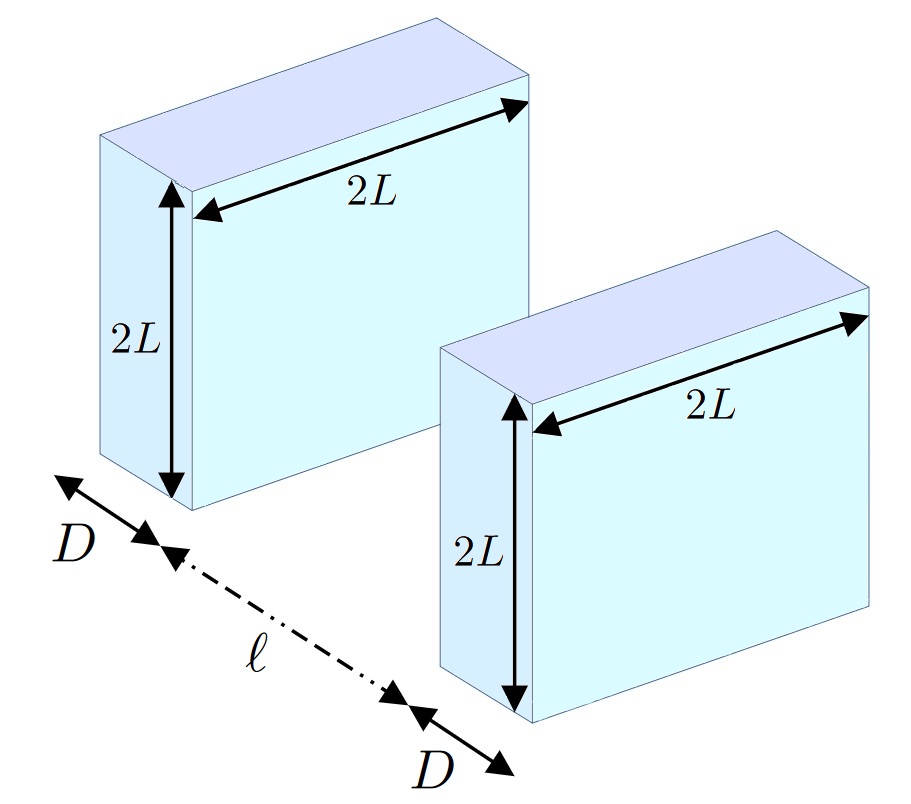}
\caption{Schematic depiction of the two parallel plates; the plates are identical in dimensions and material. We consider thin plates, which means that $D\ll\ell$. Furthermore, we take the limit $L \to \infty$. The plates are surrounded by vacuum or some kind of gas, and they themselves consist of a material with a density that leads to $D \ll 1/m_{X;\mathrm{Plate}}$ for the considered regions of the screened scalar fields models' parameter spaces.}
\label{fig:Plates}
\end{figure}

Having derived the quantum and thermal potentials between two nucleons, we can now use these results to compute the pressures between two parallel plates at separation $\ell$, with thicknesses $D$ and widths $2L$ in both transversal dimensions; see Fig.~\ref{fig:Plates}. 
The induced pressure between the two plates is computed by
\begin{eqnarray}\label{eq:pressuregen}
    \mathcal{P}^{(Z)}_{X;Y}(\ell) &=& - \frac{\partial} {\partial\ell} \frac{\mathfrak{U}^{(Z)}_{X;Y}(\ell)}{S}~,
\end{eqnarray}
where $\mathfrak{U}^{(Z)}_{X;Y}(\ell)/S$ is the integrated potential per area $S=4L^2$ of both plates with $Y \in \{Q,T \}$ and $Z \in \{1,2\}$. We restrict our discussion to the case $D \ll \ell$ and $L \to \infty$, and note that the results are only valid for parameters where $D \ll 1/m_{X;\mathrm{Plate}}$ is fulfilled. The last assumption corresponds to a field's Compton wavelength within a plate being much larger than the plate's thickness. In this case, the field cannot reach its VEV, $\varphi_{X;\mathrm{Plate}}$, within the plate but instead stays close to its VEV in vacuum\footnote{Here, the term vacuum also refers to any residual gases.}, $\varphi_{X;\mathrm{Vac}}$. Consequently, we can safely assume that $\varphi_{X;\mathrm{Plate}} \approx \varphi_{X;\mathrm{Vac}} =: \varphi_X$ and $m_{X;\mathrm{Plate}} \approx m_{X;\mathrm{Vac}} =: m_X$. In turn, the assumptions we make allow us to approximate the potential per area by 
\begin{eqnarray}
  \frac{\mathfrak{U}^{(Z)}_{X;Y}(\ell)}{S} &\approx& \lim_{L\to\infty}\frac{D^2\rho^2}{S}\int_{-L}^L dx\int_{-L}^L dx'\int_{-L}^L dy\int_{-L}^L dy' V^{(Z)}_{X;Y}\left( \sqrt{\ell^2 + (y' - y)^2 + (x' - x)^2}  \right)~,
\end{eqnarray}
where the factor $D^2$ stems from our approximation of the two integrals over the longitudinal direction, $\rho$ is the nucleon number density of each plate\footnote{We note, that the electron contributions can be safely neglected at our level of precision.}, and we set $r = \sqrt{\ell^2 + (y' - y)^2 + (x' - x)^2}$. We introduce new coordinates 
\begin{eqnarray}
  x_\pm &=& x' \pm x~, \qquad y_\pm = y' \pm y~.
\end{eqnarray}
Hence, we obtain
\begin{eqnarray}
  \frac{\mathfrak{U}^{(Z)}_{X;Y}(\ell)}{S} &\approx& \lim_{L\to\infty}\frac{D^2\rho^2}{4S}\int_{-2L}^{2L} dx_-\int_{|x_-| - 2L}^{2L - |x_-|} dx_+\int_{-2L}^{2L} dy_-\int_{|y_-| - 2L}^{2L - |y_-|} dy_+
  V^{(Z)}_{X;Y}\left( \sqrt{\ell^2 + y_-^2 + x_-^2}  \right)
  \nonumber
  \\
  &\approx& \lim_{L\to\infty}\frac{D^2\rho^2}{S}\int_{-2L}^{2L} dx_-\int_{-2L}^{2L} dy_-\left(2L - |x_-|\right)\left(2L - |y_-|\right)V^{(Z)}_{X;Y}\left( \sqrt{\ell^2 + y_-^2 + x_-^2}  \right)~.
\end{eqnarray}
Taking the limit, we find
\begin{eqnarray}
  \frac{\mathfrak{U}^{(Z)}_{X;Y}(\ell)}{S} &\approx& D^2\rho^2\int_{-\infty}^\infty dx_-\int_{-\infty}^\infty dy_- V^{(Z)}_{X;Y}\left( \sqrt{\ell^2 + y_-^2 + x_-^2}  \right)
  \nonumber
  \\
  &\approx& 2\pi D^2\rho^2 \int\limits_0^\infty d\mathfrak{r}\,\mathfrak{r} \,V^{(Z)}_{X;Y}\!\left( \sqrt{\ell^2 + \mathfrak{r}^2} \right)~.
\end{eqnarray}
Substituting $v = \sqrt{\ell^2 + \mathfrak{r}^2}$, we have 
\begin{eqnarray}
    \frac{\mathfrak{U}^{(Z)}_{X;Y}(\ell)}{S} &\approx& 2\pi  D^2\rho^2 \int\limits_\ell^\infty dv \,v\,V^{(Z)}_{X;Y}\!\left( v \right)~.
\end{eqnarray}
After substituting this into Eq.~(\ref{eq:pressuregen}), we obtain
\begin{eqnarray}
    \mathcal{P}^{(Z)}_{X;Y}(\ell) &\approx& - 2\pi D^2\rho^2 \partial_\ell  \int\limits_\ell^\infty dv \,v\,V^{(Z)}_{X;Y}\!\left( v \right)
    \nonumber
    \\
    &\approx&
     2\pi D^2\rho^2 \ell V^{(Z)}_{X;Y}\!\left( \ell \right)
    ~.
\end{eqnarray}
This means, from Eqs.~(\ref{eq:Q1}), (\ref{eq:Q2}) and (\ref{eq:T2}) we obtain:
\begin{eqnarray}
\label{eq:PQ1}
    \mathcal{P}^{(1)}_{X;Q}(\ell) 
    &\approx& -\left(g_X^{(1)}\right)^2\,\frac{ D^2\rho^2}{2}\,e^{-m_X \ell}~,
\\
\label{eq:PQ2}
    \mathcal{P}^{(2)}_{X;Q}(\ell) 
    &\approx& 
    - \left(g_X^{(2)}\right)^2\frac{m_X  D^2\rho^2 }{4\pi^2  \ell} K_1(2 m_X \ell)~,
\\
\label{eq:PT2}
    \mathcal{P}^{(2)}_{X;T}(\ell) 
    &\approx& 
    -\left(g_X^{(2)}\right)^2\frac{ D^2\rho^2}{\pi^2 } 
    \frac{T m_X}{\sqrt{1+(2\ell T)^2}} K_1\left( \frac{m_X}{T}\sqrt{1+(2\ell T)^2} \right)~.
\end{eqnarray}
Consequently, the total pressure, excluding pressures from classical fifth forces, induced by a scalar field of model $X$ is given by
\begin{eqnarray}
\label{eq:TotalPress}
   \mathcal{P}_X(\ell) &=&  \mathcal{P}^{(1)}_{X;Q}(\ell) +\mathcal{P}^{(2)}_{X;Q}(\ell) + 
   \mathcal{P}^{(2)}_{X;T}(\ell)
   \nonumber
   \\
   &\approx&
   -D^2\rho^2
   \bigg\{ 
   \left(g_X^{(1)}\right)^2
   \frac{ 1}{2}\,e^{-m_X \ell}
   + 
   \left(g_X^{(2)}\right)^2
   \frac{ m_X}{\pi^2 }
   \bigg[
   \frac{1   }{4  \ell} K_1(2 m_X \ell)
\nonumber
   \\
   &&
   ~~~~~~~~~~~~~~~~~~~~~~~~~~~~~~~~~~~~~~~~~~~~~~~~~~~~
   + 
    \frac{T }{\sqrt{1+(2\ell T)^2}} K_1\left( \frac{m_X}{T}\sqrt{1+(2\ell T)^2} \right)
    \bigg]
    \bigg\}
   ~.~~~~~~~
\end{eqnarray}
It should be noted, that each of the three pressure contributions, and as such the total pressure as well, are negative. This corresponds to an attractive force between the plates.


\subsection{Discussion}
\label{ssec:Comparisons}

\begin{figure} [htbp]
\centering
    \subfloat[][]{\includegraphics[scale=0.5]{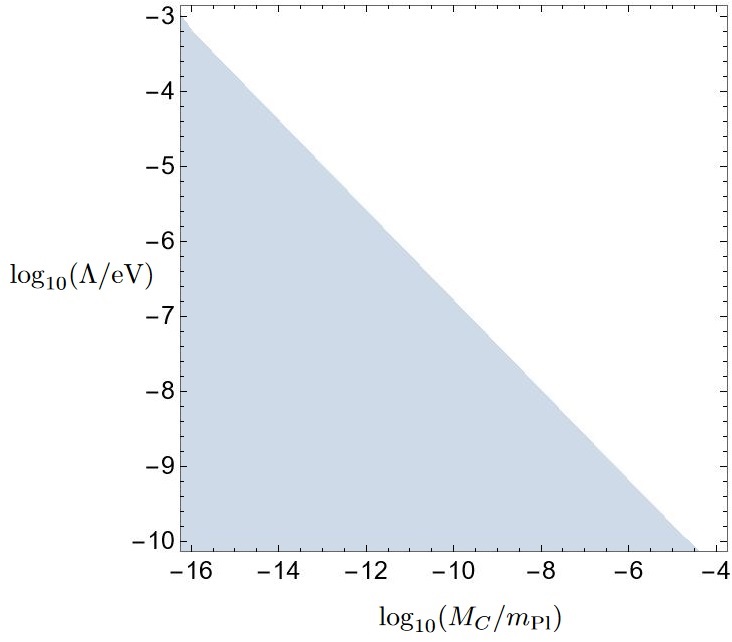}}

    \subfloat[][]{\includegraphics[scale=0.5]{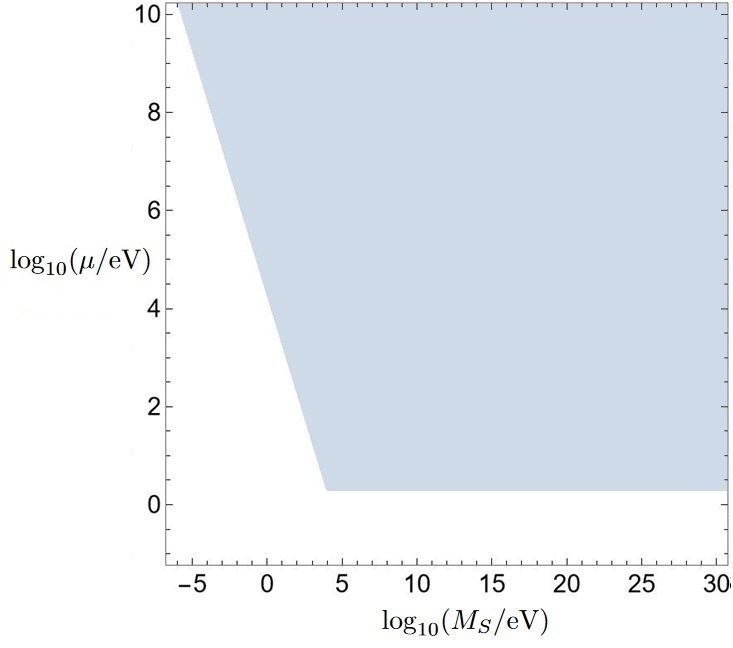}}
    \qquad
    \subfloat[][]{\includegraphics[scale=0.517]{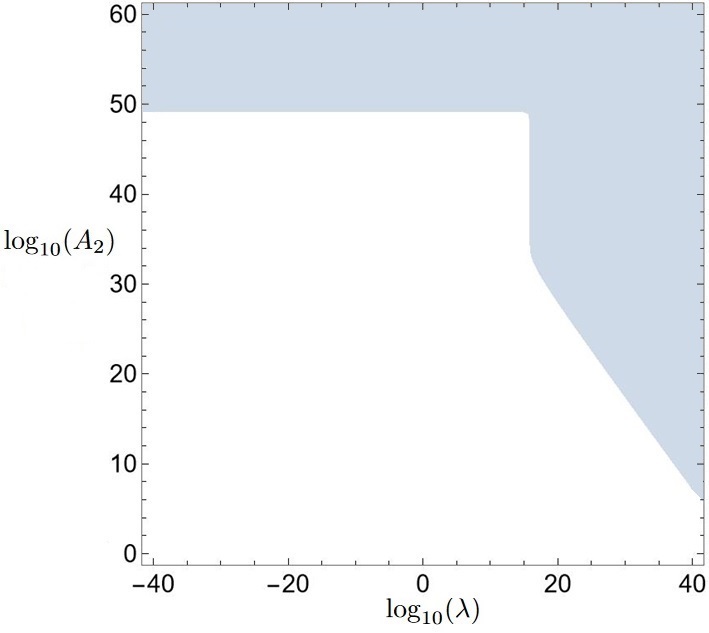}}
\caption{Areas in the model parameter spaces for which Eq.~(\ref{eq:Comparison}) is valid for a hydrogen gas with $T=300$ K and a pressure of $9.6 \times 10^{-10}$ mbar surrounding the two parallel plates; note that the plots continue where they touch one of the axes; (a): chameleon model with $n=1$; (b): symmetron in the symmetry-broken (unscreened) phase for $\lambda_S =0.1$; (c): environment-dependent dilaton with $V_0 =1 \,\text{MeV}^4$; note that the plot barely changes for larger $V_0$.}
    \label{fig:compare}
\end{figure}

We now discuss the regions of the considered screened scalar field models' parameter spaces where the induced thermal pressures are at least as large as the quantum pressures. This means that we are looking for model parameters that fulfill 
\begin{eqnarray}
\label{eq:Comparison}
   \frac{ \pi^2}{2} \left(\frac{g_X^{(1)}}{g_X^{(2)}}\right)^2\,\frac{e^{-m_X \ell}}{m_X }
    +\frac{1 }{4  \ell} K_1(2 m_X \ell)
    \leq
    \frac{T }{\sqrt{1+(2\ell T)^2}} K_1\left( \frac{m_X}{T}\sqrt{1+(2\ell T)^2} \right)~,
\end{eqnarray}
where $m_X \gg T$ due to our assumption of a Boltzmann distribution. 
In order to obtain quantitative results, we consider a plate separation of $55 \, \mu m$, as was also used in Ref.~\cite{Brax:2018grq}. In addition, we follow Ref.~\cite{Kading:2023mdk} and assume the parallel plates to be surrounded by a residual hydrogen gas of temperature $T =300$ K and with a pressure of $9.6 \times 10^{-10}$ mbar that is in thermal equilibrium with the plates and the light scalar fields. The resulting hydrogen mass density $\rho_{H_2}$ determines the VEVs and masses of the considered screened scalar field models. Using these exemplary experimental conditions, the parts of the three models' parameter spaces that fulfill Eq.~(\ref{eq:Comparison}) are depicted in Fig.~\ref{fig:compare}. 

At first, we look at the $n=1$ chameleon. Taking into account $\varphi_C\ll M_C$ and Eq.~(\ref{eq:Comparison}), we find that large parts of this model's parameter space allow for a thermal pressure at least as strong as the quantum pressure terms; see Fig.~\ref{fig:compare}(a). Interestingly, the area depicted in Fig.~\ref{fig:compare}(a) is entirely contained in the region that Ref.~\cite{Brax:2018grq} constrains by solely using the chameleon quantum pressure. This means that the thermal pressure induced by chameleons is of experimental significance and must be taken into account.

Next, we look at the symmetron model. We only consider symmetrons in the symmetry-broken phase, in which the fifth force is not screened, and which is determined by the condition $\mu^2 > \rho_{H_2}/M_S^2$. Furthermore, we set $\lambda_S=0.1$ as Ref.~\cite{Brax:2018grq} also did. In Fig.~\ref{fig:compare}(b), we see that also for large parts of the symmetron parameter space, the thermal pressure can be at least as strong as the quantum pressure. 

Finally, we look at environment-dependent dilatons. Again, we find that large parts of the model parameter space allow for a thermal pressure that is significant in comparison to the quantum pressures.
However, as of yet, nobody has computed quantum pressures for environment-dependent dilatons in specific experiments. Therefore, we cannot make a statement about the absolute magnitude of quantum and thermal pressures induced by this model. Such computations might be subject of a future work that takes into account a more realistic setup.


\section{Conclusions}
\label{sec:Conclusion}

Light scalar fields appear in many discussions throughout modern physics, and serve as candidates for dark energy or dark matter. While there are many experiments searching for evidence of such fields, there are physical aspects that are still not fully considered in experimental and theoretical analyses. A few years ago, Ref.~\cite{Brax:2018grq} showed that the pressure induced by quantum fluctuations of screened scalar fields between two infinitely wide and infinitely thick plates can be of great experimental importance. However, thermal pressures were not taken into account.

In the present article, we took first steps in the discussion of thermal pressures from light scalar fields. For this, we computed quantum and thermal potentials from one- and two-scalar exchanges between two nucleons. Next, we integrated these potentials over two thin but infinitely wide parallel plates, and subsequently derived expressions for quantum and thermal pressures from light scalar fields. Interestingly, we found that there is no thermal pressure from a single-particle exchange.

As explicit examples, we looked at three popular screened scalar field models: chameleons, symmetrons, and environment-dependent dilatons. Since we derived our general results for scalar fields with constant masses, we had to restrict our discussion to model parameters and experimental specifications for which the screened scalar fields cannot reach the minima of their potentials within each one of the two plates, such that we were allowed to set the scalar's VEVs and masses to the values corresponding to the density of the vacuum or gas the two plates are embedded in. Using an explicit experimental setup as an example, for all three considered screened scalar field models, we found large regions in their parameter spaces for which the thermal pressures are at least as significant as the quantum pressures. The region for the $n=1$ chameleon model is fully included in the area constrained in Ref.~\cite{Brax:2018grq}, which prompted us to conclude that chameleon thermal pressures are actually of experimental significance.

Our findings imply that re-evaluations of the chameleon and symmetron constraints presented in Ref.~\cite{Brax:2018grq}, that also take into account thermal pressures, might be necessary. However, this quite likely also applies to many other constraints on screened scalar fields, including environment-dependent dilatons, for which thermal effects were not taken into account. Therefore, we believe that finite-temperature effects must necessarily be discussed in future analyses of experiments that aim to constrain screened scalars or light scalar fields in general. For this, we must go beyond the initial steps taken here. This means that future analyses should make use of less assumptions than we did in ours. For example, if looking at screened scalar fields, the changes of VEVs and masses due to the presence of the two plates must be accurately described. In addition, the impact of the possible presence of a vacuum chamber or other experimental equipment should be taken into account and the assumption $D \ll \ell$ might have to be dropped. However, in this case, analytical solutions can most likely not be obtained, which necessitates the use of sophisticated numerical methods. 


\begin{acknowledgments}
This research was funded in whole or in part by the Austrian Science Fund (FWF)
\\
$[10.55776/\mathrm{P}34240]$, and $[10.55776/\mathrm{PAT}8564023]$, and is based upon work from COST Action COSMIC WISPers CA21106, supported by COST (European Cooperation in Science and Technology). For open access purposes, the author has applied a CC BY public copyright license to any author accepted manuscript version arising from this submission. The authors acknowledge TU Wien Bibliothek for financial support through its Open Access Funding Programme.
\end{acknowledgments}

\bibliography{QTP}

\providecommand{\href}[2]{#2}\begingroup\raggedright\begin{thebibliography}{10}

\bibitem{Fujii2003}
{Fujii, Yasunori and Maeda, Kei-ichi}, \emph{The Scalar-Tensor Theory of
  Gravitation}, Cambridge Monographs on Mathematical Physics. Cambridge
  University Press, 2003,
  \href{https://doi.org/10.1017/CBO9780511535093}{10.1017/CBO9780511535093}.

\bibitem{Clifton2011}
T.~Clifton, P.~G. Ferreira, A.~Padilla and C.~Skordis, \emph{Modified gravity
  and cosmology},
  \href{https://doi.org/https://doi.org/10.1016/j.physrep.2012.01.001}{\emph{Physics
  Reports} {\bfseries 513} (2012) 1}.

\bibitem{Joyce2014}
A.~Joyce, B.~Jain, J.~Khoury and M.~Trodden, \emph{{Beyond the Cosmological
  Standard Model}},
  \href{https://doi.org/10.1016/j.physrep.2014.12.002}{\emph{Phys. Rept.}
  {\bfseries 568} (2015) 1} [\href{https://arxiv.org/abs/1407.0059}{{\ttfamily
  1407.0059}}].

\bibitem{Dickey1994}
J.~O. Dickey, P.~L. Bender, J.~E. Faller, X.~X. Newhall, R.~L. Ricklefs, J.~G.
  Ries et~al., \emph{{Lunar Laser Ranging: A Continuing Legacy of the Apollo
  Program}}, \href{https://doi.org/10.1126/science.265.5171.482}{\emph{Science}
  {\bfseries 265} (1994) 482}.

\bibitem{Adelberger2003}
E.~Adelberger, B.~Heckel and A.~Nelson, \emph{{Tests of the Gravitational
  Inverse-Square Law}},
  \href{https://doi.org/10.1146/annurev.nucl.53.041002.110503}{\emph{Annual
  Review of Nuclear and Particle Science} {\bfseries 53} (2003) 77}.

\bibitem{Kapner2007}
D.~J. Kapner, T.~S. Cook, E.~G. Adelberger, J.~H. Gundlach, B.~R. Heckel, C.~D.
  Hoyle et~al., \emph{{Tests of the Gravitational Inverse-Square Law below the
  Dark-Energy Length Scale}},
  \href{https://doi.org/10.1103/PhysRevLett.98.021101}{\emph{Phys. Rev. Lett.}
  {\bfseries 98} (2007) 021101}.

\bibitem{Burrage:2016yjm}
C.~Burrage, E.~J. Copeland and P.~Millington, \emph{{Radial acceleration
  relation from symmetron fifth forces}},
  \href{https://doi.org/10.1103/PhysRevD.95.064050}{\emph{Phys. Rev. D}
  {\bfseries 95} (2017) 064050}
  [\href{https://arxiv.org/abs/1610.07529}{{\ttfamily 1610.07529}}].

\bibitem{OHare:2018ayv}
C.~A.~J. O'Hare and C.~Burrage, \emph{{Stellar kinematics from the symmetron
  fifth force in the Milky Way disk}},
  \href{https://doi.org/10.1103/PhysRevD.98.064019}{\emph{Phys. Rev. D}
  {\bfseries 98} (2018) 064019}
  [\href{https://arxiv.org/abs/1805.05226}{{\ttfamily 1805.05226}}].

\bibitem{Burrage:2018zuj}
C.~Burrage, E.~J. Copeland, C.~K\"ading and P.~Millington, \emph{{Symmetron
  scalar fields: Modified gravity, dark matter, or both?}},
  \href{https://doi.org/10.1103/PhysRevD.99.043539}{\emph{Phys. Rev. D}
  {\bfseries 99} (2019) 043539}
  [\href{https://arxiv.org/abs/1811.12301}{{\ttfamily 1811.12301}}].

\bibitem{Kading:2023hdb}
C.~K\"ading, \emph{{Lensing with Generalized Symmetrons}},
  \href{https://doi.org/10.3390/astronomy2020009}{\emph{Astronomy} {\bfseries
  2} (2023) 128} [\href{https://arxiv.org/abs/2304.05875}{{\ttfamily
  2304.05875}}].

\bibitem{Khoury2003}
J.~Khoury and A.~Weltman, \emph{{Chameleon cosmology}},
  \href{https://doi.org/10.1103/PhysRevD.69.044026}{\emph{Phys. Rev. D}
  {\bfseries 69} (2004) 044026}
  [\href{https://arxiv.org/abs/astro-ph/0309411}{{\ttfamily
  astro-ph/0309411}}].

\bibitem{Khoury20032}
J.~Khoury and A.~Weltman, \emph{{Chameleon fields: Awaiting surprises for tests
  of gravity in space}},
  \href{https://doi.org/10.1103/PhysRevLett.93.171104}{\emph{Phys. Rev. Lett.}
  {\bfseries 93} (2004) 171104}
  [\href{https://arxiv.org/abs/astro-ph/0309300}{{\ttfamily
  astro-ph/0309300}}].

\bibitem{Dehnen1992}
H.~Dehnen, H.~Frommert and F.~Ghaboussi, \emph{{Higgs field and a new scalar -
  tensor theory of gravity}},
  \href{https://doi.org/10.1007/BF00674344}{\emph{Int. J. Theor. Phys.}
  {\bfseries 31} (1992) 109}.

\bibitem{Gessner1992}
E.~Gessner, \emph{{A new scalar tensor theory for gravity and the flat rotation
  curves of spiral galaxies}},
  \href{https://doi.org/10.1007/BF00645239}{\emph{Astrophys. Space Sci.}
  {\bfseries 196} (1992) 29}.

\bibitem{Damour1994}
T.~Damour and A.~M. Polyakov, \emph{{The String dilaton and a least coupling
  principle}}, \href{https://doi.org/10.1016/0550-3213(94)90143-0}{\emph{Nucl.
  Phys. B} {\bfseries 423} (1994) 532}
  [\href{https://arxiv.org/abs/hep-th/9401069}{{\ttfamily hep-th/9401069}}].

\bibitem{Pietroni2005}
M.~Pietroni, \emph{Dark energy condensation},
  \href{https://doi.org/10.1103/PhysRevD.72.043535}{\emph{Phys. Rev. D}
  {\bfseries 72} (2005) 043535}.

\bibitem{Olive2008}
K.~A. Olive and M.~Pospelov, \emph{Environmental dependence of masses and
  coupling constants},
  \href{https://doi.org/10.1103/PhysRevD.77.043524}{\emph{Phys. Rev. D}
  {\bfseries 77} (2008) 043524}.

\bibitem{Brax:2010gi}
P.~Brax, C.~van~de Bruck, A.-C. Davis and D.~Shaw, \emph{{The Dilaton and
  Modified Gravity}},
  \href{https://doi.org/10.1103/PhysRevD.82.063519}{\emph{Phys. Rev. D}
  {\bfseries 82} (2010) 063519}
  [\href{https://arxiv.org/abs/1005.3735}{{\ttfamily 1005.3735}}].

\bibitem{Hinterbichler2010}
K.~Hinterbichler and J.~Khoury, \emph{{Symmetron Fields: Screening Long-Range
  Forces Through Local Symmetry Restoration}},
  \href{https://doi.org/10.1103/PhysRevLett.104.231301}{\emph{Phys. Rev. Lett.}
  {\bfseries 104} (2010) 231301}
  [\href{https://arxiv.org/abs/1001.4525}{{\ttfamily 1001.4525}}].

\bibitem{Hinterbichler2011}
K.~Hinterbichler, J.~Khoury, A.~Levy and A.~Matas, \emph{{Symmetron
  Cosmology}}, \href{https://doi.org/10.1103/PhysRevD.84.103521}{\emph{Phys.
  Rev. D} {\bfseries 84} (2011) 103521}
  [\href{https://arxiv.org/abs/1107.2112}{{\ttfamily 1107.2112}}].

\bibitem{Gasperini:2001pc}
M.~Gasperini, F.~Piazza and G.~Veneziano, \emph{{Quintessence as a runaway
  dilaton}}, \href{https://doi.org/10.1103/PhysRevD.65.023508}{\emph{Phys. Rev.
  D} {\bfseries 65} (2002) 023508}
  [\href{https://arxiv.org/abs/gr-qc/0108016}{{\ttfamily gr-qc/0108016}}].

\bibitem{Damour:2002nv}
T.~Damour, F.~Piazza and G.~Veneziano, \emph{{Violations of the equivalence
  principle in a dilaton runaway scenario}},
  \href{https://doi.org/10.1103/PhysRevD.66.046007}{\emph{Phys. Rev. D}
  {\bfseries 66} (2002) 046007}
  [\href{https://arxiv.org/abs/hep-th/0205111}{{\ttfamily hep-th/0205111}}].

\bibitem{Damour:2002mi}
T.~Damour, F.~Piazza and G.~Veneziano, \emph{{Runaway dilaton and equivalence
  principle violations}},
  \href{https://doi.org/10.1103/PhysRevLett.89.081601}{\emph{Phys. Rev. Lett.}
  {\bfseries 89} (2002) 081601}
  [\href{https://arxiv.org/abs/gr-qc/0204094}{{\ttfamily gr-qc/0204094}}].

\bibitem{Brax:2011ja}
P.~Brax, C.~van~de Bruck, A.-C. Davis, B.~Li and D.~J. Shaw, \emph{{Nonlinear
  Structure Formation with the Environmentally Dependent Dilaton}},
  \href{https://doi.org/10.1103/PhysRevD.83.104026}{\emph{Phys. Rev. D}
  {\bfseries 83} (2011) 104026}
  [\href{https://arxiv.org/abs/1102.3692}{{\ttfamily 1102.3692}}].

\bibitem{Brax2022}
P.~Brax, H.~Fischer, C.~K\"ading and M.~Pitschmann, \emph{{The environment
  dependent dilaton in the laboratory and the solar system}},
  \href{https://doi.org/10.1140/epjc/s10052-022-10905-w}{\emph{Eur. Phys. J. C}
  {\bfseries 82} (2022) 934}
  [\href{https://arxiv.org/abs/2203.12512}{{\ttfamily 2203.12512}}].

\bibitem{Burrage:2016bwy}
C.~Burrage and J.~Sakstein, \emph{{A Compendium of Chameleon Constraints}},
  \href{https://doi.org/10.1088/1475-7516/2016/11/045}{\emph{JCAP} {\bfseries
  11} (2016) 045} [\href{https://arxiv.org/abs/1609.01192}{{\ttfamily
  1609.01192}}].

\bibitem{Pokotilovski:2012xuk}
Y.~N. Pokotilovski, \emph{{Strongly coupled chameleon fields: Possible test
  with a neutron Lloyd's mirror interferometer}},
  \href{https://doi.org/10.1016/j.physletb.2013.01.022}{\emph{Phys. Lett. B}
  {\bfseries 719} (2013) 341}
  [\href{https://arxiv.org/abs/1203.5017}{{\ttfamily 1203.5017}}].

\bibitem{Pokotilovski:2013tma}
Y.~N. Pokotilovski, \emph{{Potential of the neutron Lloyd`s mirror
  interferometer for the search for new interactions}},
  \href{https://doi.org/10.1134/S106377611309001X}{\emph{J. Exp. Theor. Phys.}
  {\bfseries 116} (2013) 609}
  [\href{https://arxiv.org/abs/1311.4679}{{\ttfamily 1311.4679}}].

\bibitem{Jenke:2014yel}
T.~Jenke et~al., \emph{{Gravity Resonance Spectroscopy Constrains Dark Energy
  and Dark Matter Scenarios}},
  \href{https://doi.org/10.1103/PhysRevLett.112.151105}{\emph{Phys. Rev. Lett.}
  {\bfseries 112} (2014) 151105}
  [\href{https://arxiv.org/abs/1404.4099}{{\ttfamily 1404.4099}}].

\bibitem{Baum:2014rka}
S.~Baum, G.~Cantatore, D.~H.~H. Hoffmann, M.~Karuza, Y.~K. Semertzidis,
  A.~Upadhye et~al., \emph{{Detecting solar chameleons through radiation
  pressure}}, \href{https://doi.org/10.1016/j.physletb.2014.10.055}{\emph{Phys.
  Lett. B} {\bfseries 739} (2014) 167}
  [\href{https://arxiv.org/abs/1409.3852}{{\ttfamily 1409.3852}}].

\bibitem{Burrage:2014oza}
C.~Burrage, E.~J. Copeland and E.~A. Hinds, \emph{{Probing Dark Energy with
  Atom Interferometry}},
  \href{https://doi.org/10.1088/1475-7516/2015/03/042}{\emph{JCAP} {\bfseries
  03} (2015) 042} [\href{https://arxiv.org/abs/1408.1409}{{\ttfamily
  1408.1409}}].

\bibitem{Hamilton:2015zga}
P.~Hamilton, M.~Jaffe, P.~Haslinger, Q.~Simmons, H.~M\"uller and J.~Khoury,
  \emph{{Atom-interferometry constraints on dark energy}},
  \href{https://doi.org/10.1126/science.aaa8883}{\emph{Science} {\bfseries 349}
  (2015) 849} [\href{https://arxiv.org/abs/1502.03888}{{\ttfamily
  1502.03888}}].

\bibitem{Lemmel:2015kwa}
H.~Lemmel, P.~Brax, A.~N. Ivanov, T.~Jenke, G.~Pignol, M.~Pitschmann et~al.,
  \emph{{Neutron Interferometry constrains dark energy chameleon fields}},
  \href{https://doi.org/10.1016/j.physletb.2015.02.063}{\emph{Phys. Lett. B}
  {\bfseries 743} (2015) 310}
  [\href{https://arxiv.org/abs/1502.06023}{{\ttfamily 1502.06023}}].

\bibitem{Burrage:2015lya}
C.~Burrage and E.~J. Copeland, \emph{{Using Atom Interferometry to Detect Dark
  Energy}}, \href{https://doi.org/10.1080/00107514.2015.1060058}{\emph{Contemp.
  Phys.} {\bfseries 57} (2016) 164}
  [\href{https://arxiv.org/abs/1507.07493}{{\ttfamily 1507.07493}}].

\bibitem{Elder:2016yxm}
B.~Elder, J.~Khoury, P.~Haslinger, M.~Jaffe, H.~M\"uller and P.~Hamilton,
  \emph{{Chameleon Dark Energy and Atom Interferometry}},
  \href{https://doi.org/10.1103/PhysRevD.94.044051}{\emph{Phys. Rev. D}
  {\bfseries 94} (2016) 044051}
  [\href{https://arxiv.org/abs/1603.06587}{{\ttfamily 1603.06587}}].

\bibitem{Ivanov:2016rfs}
A.~N. Ivanov, G.~Cronenberg, R.~H\"ollwieser, M.~Pitschmann, T.~Jenke,
  M.~Wellenzohn et~al., \emph{{Exact solution for chameleon field, self-coupled
  through the Ratra-Peebles potential with $n=1$ and confined between two
  parallel plates}},
  \href{https://doi.org/10.1103/PhysRevD.94.085005}{\emph{Phys. Rev. D}
  {\bfseries 94} (2016) 085005}
  [\href{https://arxiv.org/abs/1606.06867}{{\ttfamily 1606.06867}}].

\bibitem{Burrage:2016rkv}
C.~Burrage, A.~Kuribayashi-Coleman, J.~Stevenson and B.~Thrussell,
  \emph{{Constraining symmetron fields with atom interferometry}},
  \href{https://doi.org/10.1088/1475-7516/2016/12/041}{\emph{JCAP} {\bfseries
  12} (2016) 041} [\href{https://arxiv.org/abs/1609.09275}{{\ttfamily
  1609.09275}}].

\bibitem{Jaffe:2016fsh}
M.~Jaffe, P.~Haslinger, V.~Xu, P.~Hamilton, A.~Upadhye, B.~Elder et~al.,
  \emph{{Author Correction: Testing sub-gravitational forces on atoms from a
  miniature in-vacuum source mass [doi: 10.1038/nphys4189]}},
  \href{https://doi.org/10.1038/s41567-023-02255-5}{\emph{Nature Phys.}
  {\bfseries 13} (2017) 938}
  [\href{https://arxiv.org/abs/1612.05171}{{\ttfamily 1612.05171}}].

\bibitem{Brax:2017hna}
P.~Brax and M.~Pitschmann, \emph{{Exact solutions to nonlinear symmetron
  theory: One- and two-mirror systems}},
  \href{https://doi.org/10.1103/PhysRevD.97.064015}{\emph{Phys. Rev. D}
  {\bfseries 97} (2018) 064015}
  [\href{https://arxiv.org/abs/1712.09852}{{\ttfamily 1712.09852}}].

\bibitem{Sabulsky:2018jma}
D.~O. Sabulsky, I.~Dutta, E.~A. Hinds, B.~Elder, C.~Burrage and E.~J. Copeland,
  \emph{{Experiment to detect dark energy forces using atom interferometry}},
  \href{https://doi.org/10.1103/PhysRevLett.123.061102}{\emph{Phys. Rev. Lett.}
  {\bfseries 123} (2019) 061102}
  [\href{https://arxiv.org/abs/1812.08244}{{\ttfamily 1812.08244}}].

\bibitem{Brax:2018iyo}
P.~Brax, C.~Burrage and A.-C. Davis, \emph{{Laboratory constraints}},
  \href{https://doi.org/10.1142/S0218271818480097}{\emph{Int. J. Mod. Phys. D}
  {\bfseries 27} (2018) 1848009}.

\bibitem{Cronenberg:2018qxf}
G.~Cronenberg, P.~Brax, H.~Filter, P.~Geltenbort, T.~Jenke, G.~Pignol et~al.,
  \emph{{Acoustic Rabi oscillations between gravitational quantum states and
  impact on symmetron dark energy}},
  \href{https://doi.org/10.1038/s41567-018-0205-x}{\emph{Nature Phys.}
  {\bfseries 14} (2018) 1022}
  [\href{https://arxiv.org/abs/1902.08775}{{\ttfamily 1902.08775}}].

\bibitem{Hartley:2018lvm}
D.~Hartley, C.~K\"ading, R.~Howl and I.~Fuentes, \emph{{Quantum simulation of
  dark energy candidates}},
  \href{https://doi.org/10.1103/PhysRevD.99.105002}{\emph{Phys. Rev. D}
  {\bfseries 99} (2019) 105002}
  [\href{https://arxiv.org/abs/1811.06927}{{\ttfamily 1811.06927}}].

\bibitem{Zhang:2019hos}
X.~Zhang, R.~Niu and W.~Zhao, \emph{{Constraining the scalar-tensor gravity
  theories with and without screening mechanisms by combined observations}},
  \href{https://doi.org/10.1103/PhysRevD.100.024038}{\emph{Phys. Rev. D}
  {\bfseries 100} (2019) 024038}
  [\href{https://arxiv.org/abs/1906.10791}{{\ttfamily 1906.10791}}].

\bibitem{ArguedasCuendis:2019fxj}
S.~Arguedas~Cuendis et~al., \emph{{First Results on the Search for Chameleons
  with the KWISP Detector at CAST}},
  \href{https://doi.org/10.1016/j.dark.2019.100367}{\emph{Phys. Dark Univ.}
  {\bfseries 26} (2019) 100367}
  [\href{https://arxiv.org/abs/1906.01084}{{\ttfamily 1906.01084}}].

\bibitem{Jenke:2020obe}
T.~Jenke, J.~Bosina, J.~Micko, M.~Pitschmann, R.~Sedmik and H.~Abele,
  \emph{{Gravity resonance spectroscopy and dark energy symmetron fields:
  qBOUNCE experiments performed with Rabi and Ramsey spectroscopy}},
  \href{https://doi.org/10.1140/epjs/s11734-021-00088-y}{\emph{Eur. Phys. J.
  ST} {\bfseries 230} (2021) 1131}
  [\href{https://arxiv.org/abs/2012.07472}{{\ttfamily 2012.07472}}].

\bibitem{Pitschmann:2020ejb}
M.~Pitschmann, \emph{{Exact solutions to nonlinear symmetron theory: One- and
  two-mirror systems. II.}},
  \href{https://doi.org/10.1103/PhysRevD.103.084013}{\emph{Phys. Rev. D}
  {\bfseries 103} (2021) 084013}
  [\href{https://arxiv.org/abs/2012.12752}{{\ttfamily 2012.12752}}].

\bibitem{Brax:2021wcv}
P.~Brax, S.~Casas, H.~Desmond and B.~Elder, \emph{{Testing Screened Modified
  Gravity}}, \href{https://doi.org/10.3390/universe8010011}{\emph{Universe}
  {\bfseries 8} (2021) 11} [\href{https://arxiv.org/abs/2201.10817}{{\ttfamily
  2201.10817}}].

\bibitem{Qvarfort:2021zrl}
S.~Qvarfort, D.~R\"atzel and S.~Stopyra, \emph{{Constraining modified gravity
  with quantum optomechanics}},
  \href{https://doi.org/10.1088/1367-2630/ac3e1b}{\emph{New J. Phys.}
  {\bfseries 24} (2022) 033009}
  [\href{https://arxiv.org/abs/2108.00742}{{\ttfamily 2108.00742}}].

\bibitem{Yin:2022geb}
P.~Yin, R.~Li, C.~Yin, X.~Xu, X.~Bian, H.~Xie et~al., \emph{{Experiments with
  levitated force sensor challenge theories of dark energy}},
  \href{https://doi.org/10.1038/s41567-022-01706-9}{\emph{Nature Phys.}
  {\bfseries 18} (2022) 1181}.

\bibitem{Betz:2022djh}
J.~Betz, J.~Manley, E.~M. Wright, D.~Grin and S.~Singh, \emph{{Searching for
  Chameleon Dark Energy with Mechanical Systems}},
  \href{https://doi.org/10.1103/PhysRevLett.129.131302}{\emph{Phys. Rev. Lett.}
  {\bfseries 129} (2022) 131302}
  [\href{https://arxiv.org/abs/2201.12372}{{\ttfamily 2201.12372}}].

\bibitem{Brax:2022olf}
P.~Brax, A.-C. Davis and B.~Elder, \emph{{Screened scalar fields in hydrogen
  and muonium}}, \href{https://doi.org/10.1103/PhysRevD.107.044008}{\emph{Phys.
  Rev. D} {\bfseries 107} (2023) 044008}
  [\href{https://arxiv.org/abs/2207.11633}{{\ttfamily 2207.11633}}].

\bibitem{Hartley:2019wzu}
D.~Hartley, C.~K\"ading, R.~Howl and I.~Fuentes, \emph{{Quantum-enhanced
  screened dark energy detection}},
  \href{https://doi.org/10.1140/epjc/s10052-023-12360-7}{\emph{Eur. Phys. J. C}
  {\bfseries 84} (2024) 49} [\href{https://arxiv.org/abs/1909.02272}{{\ttfamily
  1909.02272}}].

\bibitem{Fischer:2023koa}
H.~Fischer, C.~K\"ading, R.~I.~P. Sedmik, H.~Abele, P.~Brax and M.~Pitschmann,
  \emph{{Search for environment-dependent dilatons}},
  \href{https://doi.org/10.1016/j.dark.2024.101419}{\emph{Phys. Dark Univ.}
  {\bfseries 43} (2024) 101419}
  [\href{https://arxiv.org/abs/2307.00243}{{\ttfamily 2307.00243}}].

\bibitem{Fischer:2023eww}
H.~Fischer, C.~K\"ading, H.~Lemmel, S.~Sponar and M.~Pitschmann, \emph{{Search
  for dark energy with neutron interferometry}},
  \href{https://doi.org/10.1093/ptep/ptae014}{\emph{PTEP} {\bfseries 2024}
  (2024) 023E02} [\href{https://arxiv.org/abs/2310.18109}{{\ttfamily
  2310.18109}}].

\bibitem{Fischer:2024coj}
H.~Fischer and R.~I.~P. Sedmik, \emph{{Numerical methods for scalar field dark
  energy in tabletop experiments and Lunar Laser Ranging}},
  \href{https://doi.org/10.1088/1475-7516/2024/10/026}{\emph{JCAP} {\bfseries
  10} (2024) 026} [\href{https://arxiv.org/abs/2401.16179}{{\ttfamily
  2401.16179}}].

\bibitem{Klimchitskaya:2024dvk}
G.~L. Klimchitskaya and V.~M. Mostepanenko, \emph{{The Nature of Dark Energy
  and Constraints on Its Hypothetical Constituents from Force Measurements}},
  \href{https://doi.org/10.3390/universe10030119}{\emph{Universe} {\bfseries
  10} (2024) 119} [\href{https://arxiv.org/abs/2403.05988}{{\ttfamily
  2403.05988}}].

\bibitem{Rene}
H.~Haghmoradi, H.~Fischer, A.~Bertolini, I.~Gali\'c, F.~Intravaia,
  M.~Pitschmann et~al., \emph{{Force metrology with plane parallel plates:
  Final design review and outlook}}, {\emph{MDPI Physics} {\bfseries 6} (2024)
  } [\href{https://arxiv.org/abs/2403.10998}{{\ttfamily 2403.10998}}].

\bibitem{Baez-Camargo:2024jia}
A.~L. B\'aez-Camargo, D.~Hartley, C.~K\"ading and I.~Fuentes, \emph{{Dynamical
  Casimir effect with screened scalar fields}},
  \href{https://doi.org/10.1116/5.0222082}{\emph{AVS Quantum Sci.} {\bfseries
  6} (2024) 045001} [\href{https://arxiv.org/abs/2404.02630}{{\ttfamily
  2404.02630}}].

\bibitem{OShea:2024jjw}
T.~O'Shea, A.-C. Davis, M.~Giannotti, S.~Vagnozzi, L.~Visinelli and J.~K.
  Vogel, \emph{{Solar chameleons: Novel channels}},
  \href{https://doi.org/10.1103/PhysRevD.110.063027}{\emph{Phys. Rev. D}
  {\bfseries 110} (2024) 063027}
  [\href{https://arxiv.org/abs/2406.01691}{{\ttfamily 2406.01691}}].

\bibitem{Mueller2024}
C.~D. Panda, M.~J. Trao, M.~Ceja, J.~Khoury, G.~M. Tino and H.~M\"uller,
  \emph{{Measuring gravitational attraction with a lattice atom
  interferometer}},
  \href{https://doi.org/10.1038/s41586-024-07561-3}{\emph{Nature} (2024) }
  [\href{https://arxiv.org/abs/2310.01344}{{\ttfamily 2310.01344}}].

\bibitem{Fischer:2024eic}
H.~Fischer, C.~K\"ading and M.~Pitschmann, \emph{{Screened Scalar Fields in the
  Laboratory and the Solar System}},
  \href{https://doi.org/10.3390/universe10070297}{\emph{Universe} {\bfseries
  10} (2024) 297} [\href{https://arxiv.org/abs/2405.14638}{{\ttfamily
  2405.14638}}].

\bibitem{Burrage:2017qrf}
C.~Burrage and J.~Sakstein, \emph{{Tests of Chameleon Gravity}},
  \href{https://doi.org/10.1007/s41114-018-0011-x}{\emph{Living Rev. Rel.}
  {\bfseries 21} (2018) 1} [\href{https://arxiv.org/abs/1709.09071}{{\ttfamily
  1709.09071}}].

\bibitem{Brax:2018grq}
P.~Brax and S.~Fichet, \emph{{Quantum Chameleons}},
  \href{https://doi.org/10.1103/PhysRevD.99.104049}{\emph{Phys. Rev. D}
  {\bfseries 99} (2019) 104049}
  [\href{https://arxiv.org/abs/1809.10166}{{\ttfamily 1809.10166}}].

\bibitem{Burrage:2018pyg}
C.~Burrage, C.~K\"ading, P.~Millington and J.~Min\'a\v{r}, \emph{{Open quantum
  dynamics induced by light scalar fields}},
  \href{https://doi.org/10.1103/PhysRevD.100.076003}{\emph{Phys. Rev. D}
  {\bfseries 100} (2019) 076003}
  [\href{https://arxiv.org/abs/1812.08760}{{\ttfamily 1812.08760}}].

\bibitem{Burrage:2019szw}
C.~Burrage, C.~K\"ading, P.~Millington and J.~Min\'a\v{r}, \emph{{Influence
  functionals, decoherence and conformally coupled scalars}},
  \href{https://doi.org/10.1088/1742-6596/1275/1/012041}{\emph{J. Phys. Conf.
  Ser.} {\bfseries 1275} (2019) 012041}
  [\href{https://arxiv.org/abs/1902.09607}{{\ttfamily 1902.09607}}].

\bibitem{Kading:2023mdk}
C.~K\"ading, M.~Pitschmann and C.~Voith, \emph{{Dilaton-induced open quantum
  dynamics}}, \href{https://doi.org/10.1140/epjc/s10052-023-11939-4}{\emph{Eur.
  Phys. J. C} {\bfseries 83} (2023) 767}
  [\href{https://arxiv.org/abs/2306.10896}{{\ttfamily 2306.10896}}].

\bibitem{SevillanoMunoz:2024ayh}
S.~Sevillano Mu\~noz, \emph{{A particle's perspective on screening
  mechanisms}},  \href{https://arxiv.org/abs/2407.08779}{{\ttfamily
  2407.08779}}.

\bibitem{Sedmik:2021iaw}
R.~I.~P. Sedmik and M.~Pitschmann, \emph{{Next Generation Design and Prospects
  for Cannex}}, \href{https://doi.org/10.3390/universe7070234}{\emph{Universe}
  {\bfseries 7} (2021) 234} [\href{https://arxiv.org/abs/2107.07645}{{\ttfamily
  2107.07645}}].

\bibitem{Brax:2014zta}
P.~Brax and A.-C. Davis, \emph{{Casimir, Gravitational and Neutron Tests of
  Dark Energy}}, \href{https://doi.org/10.1103/PhysRevD.91.063503}{\emph{Phys.
  Rev. D} {\bfseries 91} (2015) 063503}
  [\href{https://arxiv.org/abs/1412.2080}{{\ttfamily 1412.2080}}].

\bibitem{Almasi:2015zpa}
A.~Almasi, P.~Brax, D.~Iannuzzi and R.~I.~P. Sedmik, \emph{{Force sensor for
  chameleon and Casimir force experiments with parallel-plate configuration}},
  \href{https://doi.org/10.1103/PhysRevD.91.102002}{\emph{Phys. Rev. D}
  {\bfseries 91} (2015) 102002}
  [\href{https://arxiv.org/abs/1505.01763}{{\ttfamily 1505.01763}}].

\bibitem{Elder:2019yyp}
B.~Elder, V.~Vardanyan, Y.~Akrami, P.~Brax, A.-C. Davis and R.~S. Decca,
  \emph{{Classical symmetron force in Casimir experiments}},
  \href{https://doi.org/10.1103/PhysRevD.101.064065}{\emph{Phys. Rev. D}
  {\bfseries 101} (2020) 064065}
  [\href{https://arxiv.org/abs/1912.10015}{{\ttfamily 1912.10015}}].

\bibitem{Brax:2022uiv}
P.~Brax, A.-C. Davis and B.~Elder, \emph{{Casimir tests of scalar-tensor
  theories}}, \href{https://doi.org/10.1103/PhysRevD.107.084025}{\emph{Phys.
  Rev. D} {\bfseries 107} (2023) 084025}
  [\href{https://arxiv.org/abs/2211.07840}{{\ttfamily 2211.07840}}].

\bibitem{MarioHabil}
M.~Pitschmann, \emph{{The High Precision Frontier: Search for New Physics with
  “Tabletop Experiments” \& Beyond}}, habilitation, TU Wien, 2023.

\bibitem{Ferrer:1998rw}
F.~Ferrer and M.~Nowakowski, \emph{{Higgs and Goldstone bosons mediated long
  range forces}}, \href{https://doi.org/10.1103/PhysRevD.59.075009}{\emph{Phys.
  Rev. D} {\bfseries 59} (1999) 075009}
  [\href{https://arxiv.org/abs/hep-ph/9810550}{{\ttfamily hep-ph/9810550}}].

\bibitem{Feinberg:1989ps}
G.~Feinberg, J.~Sucher and C.~K. Au, \emph{{The Dispersion Theory of Dispersion
  Forces}}, \href{https://doi.org/10.1016/0370-1573(89)90111-7}{\emph{Phys.
  Rept.} {\bfseries 180} (1989) 83}.

\bibitem{Ferrer:1998ju}
F.~Ferrer, J.~A. Grifols and M.~Nowakowski, \emph{{Long range forces induced by
  neutrinos at finite temperature}},
  \href{https://doi.org/10.1016/S0370-2693(98)01489-0}{\emph{Phys. Lett. B}
  {\bfseries 446} (1999) 111}
  [\href{https://arxiv.org/abs/hep-ph/9806438}{{\ttfamily hep-ph/9806438}}].

\bibitem{Horowitz:1993kw}
C.~J. Horowitz and J.~T. Pantaleone, \emph{{Long range forces from the
  cosmological neutrinos background}},
  \href{https://doi.org/10.1016/0370-2693(93)90800-W}{\emph{Phys. Lett. B}
  {\bfseries 319} (1993) 186}
  [\href{https://arxiv.org/abs/hep-ph/9306222}{{\ttfamily hep-ph/9306222}}].

\end{thebibliography}\endgroup
\bibliographystyle{JHEP}

\end{document}